\documentstyle[aps]{revtex}

\begin{document}
\title{On the absence of dissipative instability of negative energy waves 
in hydrodynamic shear flows.}
\author{S. Chatterjee and P. S. Joarder\\
{\it Indian Institute of Astrophysics,\\
Koramangala, Bangalore -- 560 034, India.}}

\maketitle

\begin{abstract}
\noindent Stability criterion for the {\it surface gravity capillary waves} in a flowing
two-layered fluid system with viscous dissipation is investigated. It is seen
that the {\it dissipative instability} of negative energy waves is absent,- 
contrary to what earlier authors have concluded. Their error is identified to
arise from an erroneous choice of the dissipation law, in which the wave profile
velocity is wrongly equated to the particle velocity. Our corrected dissipation
law is also shown to restore {\it Galilean invariance} to the stability 
condition of the system. 
\end{abstract}

\vspace{1.0cm}



\noindent Dissipative instability in a flowing system has been discussed in the 
literature, and is being sought to be applied to explain various phenomena for 
a few decades. This instability was shown by many  authors to be caused by 
viscosity, but seen only in some selected frames of reference in which a 
particular mode of the system possesses {\it negetaive energy} 
\cite{lan,ben,kik,wei,cai,ostst,cra,oset,ryu,rudg,ruet}. Their result, 
therefore, implies that  the stability of such a system is frame dependent, 
which obviously violates the condition of {\it Galilean invariance}. In the 
present paper, it is shown that the violation of Galilean invariance arose due 
to a wrong choice of the dissipation law. This question has also been recently 
addressed by us in an earlier paper \cite{chat}, where we calculated the total 
energy in a {\it magnetohydrodynamic shear flow}. It is there shown that the 
existing theories give a rate of entropy production which is not invariant under 
Galilean transformation. The method to calculate the correct rate of energy 
dissipation is given by us in the earlier paper in considerable detail. In this 
research note, we analyse the problem from the point of view of Euler's equation 
of motion and show the validity of our earlier conclusions.

The system considered is a single interface of discontinuity separating two 
uniform {\it incompressible} fluid  media. The interface is along the $x$-
direction with the force of gravity acting downwards, i.e., in the negative $z$-
direction. The equilibrium pressures, densities and velocities are given by

\begin{eqnarray}
 p_{\rm 0}(z),~ \rho_{\rm 0}(z), ~ u_{\rm 0}(z) & = &\left\{ \begin{array}{ll}
p_{1},~\rho_{1},~u_{1},    & \mbox{$z\leq \rm{0},$} \\
p_{2},~\rho_{2},~u_{2},    & \mbox{$z> \rm{0},$} 
\end{array}
 \right . 
\end{eqnarray}

\noindent with the equlibrium pressure balance condition requiring that 
$p_{\rm 1}=p_{2}$.

We consider small perturbations about the above equilibrium configuration. 
The vertical displacement 
$\eta(x,z,t)$ of a streamline at any point $(x,z)$ can then be written in a form

\begin{eqnarray}
  \eta(x,z,t) & \equiv &  \left\{ \begin{array}{ll}
              \tilde{\eta}(k,t)\exp{\left(ikx\right)}\exp{\left\{
kz\right\}},  & \mbox{$z\leq 0$,}\\
              \tilde{\eta}(k,t)\exp{\left(ikx\right)}\exp{\left\{
- kz\right\}},   &\mbox{$z>  0$,}
        \end{array}
        \right. 
\end{eqnarray}

\noindent that is consistent with an incompressible and irrotational flow. From 
the {\it linearised equations of hydrodynamics}, we then obtain the expressions
for the velocity and the pressure fluctuations. These expressions are (with 
$i=$1,2)

\newcounter{lett1}
\setcounter{lett1}{3}
\setcounter{equation}{0}
\renewcommand{\theequation}{\arabic{lett1} \alph{equation}}

\begin{eqnarray}
\delta u_{xi}& = &\mp \left(n+ku_{i}\right)\tilde{\eta}(k,t)\exp{
\left(ikx\right)}\exp{\left\{\pm (kz)\right\}}, \\
& & \nonumber \\
\delta u_{zi}& = &\left(\frac{n+ku_{i}}{n}\right)\dot{\tilde{\eta}}(k,t)\exp{
\left(ikx\right)}\exp{\left\{\pm (kz)\right\}}, 
\end{eqnarray}

\noindent and 
\vspace{-1.0cm}

\begin{eqnarray}
& & \nonumber \\
\delta p_{i}& = &\pm\left(\frac{\rho_{i}}{k}\right)\left(n+ku_{i}\right)^2
\tilde{\eta}(k,t)\exp{ \left(ikx\right)}\exp{\left\{\pm (kz)\right\}}, 
\end{eqnarray}

\setcounter{equation}{3}
\renewcommand{\theequation}{\arabic{equation}}

\noindent with the upper sign designating $i=$ 1 and the lower sign 
designating $i=$ 2. In Eq.(3), we have assumed the temporal dependence 
to be $\tilde{\eta}(k,t)\sim \exp{\left(int\right)}$, with a `dot' designating
a time derivative $\partial/\partial t$. 

At the interface ($z=0$), the continuity of pressure fluctuations demand

\begin{eqnarray}
\delta p_{\rm 1}(x,t)- \delta p_{\rm 2}(x,t)& = & - T \frac{\partial^2}
{\partial x^2} \eta(x,t) + g\left(\rho_{\rm 1}-\rho_{\rm 2}\right)\eta(x,t),
\end{eqnarray}

\noindent with $g$ being the acceleration due to gravity, and $T$ being the suface 
tension of the interface. While solving for the temporal Fourier amplitudes 
(i.e., the normal modes) $\overline{\eta}(k,n)$ by substituing (3) in (4), we 
find that for a non-trivial solution to exist, one must have

\begin{eqnarray}
\left(n+k\overline{U}\right)^{2}&=& \left(\Delta n\right)^{2},
\end{eqnarray}

\noindent where,

\begin{eqnarray}
\left(\Delta n\right)^{2} &=& gk\left(\alpha_{\rm 1}-\alpha_{\rm 2}\right)
+k^{\rm 3}T^{\prime}-k^{\rm 2}\alpha_{\rm 1}\alpha_{\rm 2}\left(u_{\rm 1}-
u_{\rm 2}\right)^{\rm 2},
\end{eqnarray}

\noindent with

\newcounter{lett2}
\setcounter{lett2}{7}
\setcounter{equation}{0}
\renewcommand{\theequation}{\arabic{lett2} \alph{equation}}

\begin{eqnarray} 
& & \nonumber \\
\alpha_{i} &=& \rho_{i}/\left(\rho_{\rm 1}+\rho_{\rm 2}\right), (i=1,2),\\
& & \nonumber \\ 
\overline{U} &=& \left(\alpha_{\rm 1}u_{\rm 1}+\alpha_{\rm 2}u_{\rm 2}\right),
\end{eqnarray}

\noindent and 
\vspace{-1.0cm}

\begin{eqnarray}
& & \nonumber \\
T^{\prime} &=& T/\left(\rho_{\rm 1}+\rho_{\rm 2}\right).
\end{eqnarray}

\setcounter{equation}{7}
\renewcommand{\theequation}{\arabic{equation}}

\noindent Eqs.(5-7) constitute the familiar dispersion relation 
for the {\it surface gravity capillary waves} that exihibits the 
{\it Rayleigh-Taylor} and the {\it Kelvin-Helmholtz instabilities} under certain 
conditions \cite{chandra}.

It is necessary at this stage to introduce the question of dissipative 
instability. It was argued by several authors 
\cite{lan,ben,kik,wei,cai,ostst,cra,oset,ryu,rudg,ruet} 
that, on introducing a kinematic viscosity $\nu$ in any one of the two media 
(in the lower medium, say), one changes the dispersion relation to 

\begin{eqnarray}
\left(n+k\overline{U}\right)^{2} &=& \left(\Delta n\right)^{2}+i\nu
\alpha_{1}nk^{2},
\end{eqnarray}

\noindent which, for $(\Delta n)^{2}>0$ and for a small kinematic viscosity 
$\left(\nu \alpha_{1}k^{3}\overline{U}/(\Delta n)^2 \ll 1\right)$, gives the two 
roots as

\begin{eqnarray}
n_{\pm} &=& -k\left(\overline{U}\pm (\Delta n)/k\right) + \frac{i\nu\alpha_{1}k^2}
{2(\Delta n)}\left[(\Delta n)\pm k\overline{U}\right].
\end{eqnarray}

\noindent For $\overline{U}> (\Delta n)/k$, the (-) root in Eq.(9) is 
called a {\it negative energy wave} that grows in time as $\exp{\left[\nu
\alpha_{1}k^2\left(k\overline{U}/(\Delta n)-1\right)/2\right]}$, thus giving 
{\it dissipative instability} of the negative energy wave. Note that, the growth 
is possible for any non-zero but small value of $\nu$ whenever $\overline{U}> 
(\Delta n)/k$, while for $\nu$ exactly equal to zero, the instability criterion 
has no dependence on $\overline{U}$ and is given by \cite{chandra} $(\Delta n)^2 
<0$, or, 

\begin{eqnarray}
k^2\alpha_{1}\alpha_{2}\left(u_{1}-u_{2}\right)^2 > gk\left\{\left(\alpha_{1}
-\alpha_{2}\right) +k^2T^{\prime}/g\right\}.
\end{eqnarray}

The above result contains two surprising conclusions. Firstly, the response
of the system is not continuous with respect to $\nu$ as $\nu \longrightarrow 0$,
i.e., the stability of the system for an arbitrarily small viscosity is 
different from that when viscosity is exactly zero. Furthermore, the stability 
with a small but nonzero $\nu$ appears to be dependent on $\overline{U}$, where
 $\overline{U}$ is of course dependent on the frame of reference. This means, 
that the stability of the system depends on the frame of reference of the 
observer. In other words, by moving the observer with a given speed, one can 
create an instability of the negative energy waves,- a result which obviously 
violates the fundamental law of {\it Galilean invariance}.

The problem can, however, be resolved in the following way. As has been done by
most authors, we consider one of the fluids, i.e., the lower one to be viscous,
while the upper one to be non-viscous. This simplification enables us to ignore
the complications due to boundary layers, while the essential physics remains
unaltered. We note, that by substituting $in \equiv \partial/\partial t$ and
$ik\equiv \partial/\partial x$ in the dispersion relation (5) for the non-viscous
case, we obtain an equation of motion

\begin{eqnarray}
\frac{D^{2}}{Dt^{2}}\eta(x,t) &=&- (\Delta n)^{2}\eta(x,t),
\end{eqnarray}

\noindent where, $D/Dt$ is the total derivative

\begin{eqnarray}
\frac{D}{Dt}\equiv \frac{\partial}{\partial t} + \overline{U}\frac{\partial}
{\partial x}.
\end{eqnarray}

From Eq.(12), it is seen that, Eq.(11) describes the force equation 
of a fluid system moving with a nett velocity $\overline{U}$ with respect to 
the given frame of reference. Due cognizance should be taken about this fact 
while calculating the viscous force term $\nu\rho_{\rm 1}\partial^{\rm 2}/
\partial x^{\rm 2} (\delta v_{\rm z})$. It is to be noted that the term 
$\delta v_{\rm z}$ here denotes the {\it real velocity of the fluid particles} 
pertaining to wave propagation \cite{land}, and not the profile velocity 
$\partial\eta(x,t)/ \partial t$, as the earlier authors have suggested. 
We note that, in this moving fluid system, the particle velocity is calculated
as $\delta v_{\rm z}= D\eta(x,t)/Dt = i(n+k\overline{U})\eta(x,t)$.

The above argument implies that, the equation of motion in the presence of 
viscosity must read

\begin{eqnarray}
\frac{D^2}{Dt^2}\tilde{\eta}(k,t) &=& - (\Delta n)^2\tilde{\eta}(k,t) - \nu\alpha_{1}k^2\frac{D}
{Dt}\tilde{\eta}(k,t),
\end{eqnarray}

\noindent thus giving a dispersion relation

\begin{eqnarray}
\left(n+k\overline{U}\right)^2 &=&  (\Delta n)^2 + i\nu\alpha_{1}\left(n+k
\overline{U}\right)k^2,
\end{eqnarray}

\noindent that yields the two roots

\begin{eqnarray}
n_{\pm} & = & - k\left[\overline{U}\pm \frac{1}{k}\left\{(\Delta n)^2 - {\nu}^2
\alpha_{1}^2k^4/4\right\}^{1/2}\right] + i\nu\alpha_{1}k^2/2.
\end{eqnarray}

In Eq.(15), the last term on the right hand side predicts a damping 
for both the wavemodes when $(\Delta n)^{2}>\nu^2\alpha_{1}^2k^{4}/4$. A growth 
is, however, possible if and only if

\begin{eqnarray}
(\Delta n)^2 - \nu^{2}\alpha_{1}^2k^{4}/4 < 0, {\mbox{\rm and}}  \left\{\nu^{2}
\alpha_{1}^2k^{4}/4 - (\Delta n)^{2}\right\}^{1/2} > \nu\alpha_{1}k^{2}/2,
\end{eqnarray}

\noindent thus presenting the same instability criterion $(\Delta n)^{2} <0$,
 as in equation (10) for the non-viscous case. While precluding the dissipative 
instability of negative energy waves, Eqs.(15) and (16) thus suggest that the 
presence of viscous dissipation does not at all alter the stability property 
of the surface gravity capillary waves.

The foregoing analysis shows that, the dissipative instability is simply an
artifact of an erroneous choice of the viscous damping law by the earlier
authors. The correct viscosity law, written as in the last term of Eq.(13),
corresponds to a resistance proportional to $\nu\rho_{1}\partial^2
(\delta v_{\rm z})/\partial x^2 = -i\nu\rho_{1}k^2\left(n+k\overline{U}
\right)\eta(x,t)$. Replacing this expression by $-i\nu\rho_{1}nk^2\eta(x,t)$ 
, as was done by the earlier authors, would be equivalent to having a 
resistance of the form $\alpha\left(n,\overline{U}\right)\nu\rho_{1}
\partial^2(\delta v_{\rm z})/\partial x^2$, with $\alpha(n,\overline{U})= 
n/\left(n+k\overline{U}\right)$. For small values of the kinematic viscosity 
$\nu$, we can then use Eq.(5) to write $\alpha(n_{+},\overline{U})
\approx \left(k\overline{U}+\Delta n\right)/\Delta n$ and $\alpha(n_{-},
\overline{U})\approx \left(-k\overline{U}+\Delta n\right)/\Delta n$ for the
(+) and the (-) modes, respectively. Here, $\alpha(n_{+},\overline{U})$ 
is always positive, but $\alpha(n_{-},\overline{U})$ is negative when 
$\overline{U}> (\Delta n)/k$.  In such a situation, the force $\alpha(n_{-},
\overline{U})\nu\rho_{1}\partial^2 (\delta v_{\rm z})/\partial x^2$ would 
act as an attractive force, that helps to build up the amplitude of the 
negative energy wave. The erroneous resistance formula, that has been used 
so far in the literature, thus makes the viscous drag force frame dependent 
and gives a velocity dependent acceleration in selected frames, rather than 
a deceleration in all frames. By substituting unity for $\alpha(n,
\overline{U})$, the correct viscous resistance formula however demands 
that, the velocity $\delta v_{\rm z}$ be the true velocity of the fluid 
particles pertaing to wave motion so that, the viscous force becomes frame 
independent. The use of this correct formula leads us to two important 
results. Firstly, it precludes the possibility of dissipative instability 
in negative energy wave systems. Secondly, it gives a dispersion relation 
in which the drift velocity $\overline{U}$ appears only as a Doppler shift 
term in the frequency;- the important consequence of which is that, the 
stability of the system is independent of $\overline{U}$, thus giving the 
required frame independent universal character to the stability condition
of the system.

\section*{Acknowledgements.}

\noindent PSJ is grateful to the solar theory group of the University of St.
Andrews, and particularly to Prof. B. Roberts for offering him a PPARC visiting
fellowship in the university. He also thanks Dr. V. Nakariakov, Prof. R. A. Cairns 
and Prof. A. D. D. Craik for stimulating his interest in negative energy waves.
We wish to thank Profs. M. H. Gokhale, S. S. Hasan, C. Uberoi and P. Venkatakrishnan
for discussions.

\end{document}